# Atmospheric pressure graphitization of SiC(0001) – A route towards wafer-size graphene layers

Konstantin V. Emtsev[1], Aaron Bostwick[2], Karsten Horn[3], Johannes Jobst[4], Gary L. Kellogg[5], Lothar Ley[1], Jessica L. McChesney[2], Taisuke Ohta[5], Sergey A. Reshanov[4], Eli Rotenberg[2], Andreas K. Schmid[6], Daniel Waldmann[4], Heiko B. Weber[4], Thomas Seyller[1,*]

[1] Lehrstuhl für Technische Physik, Friedrich-Alexander-Universität Erlangen-Nürnberg, Erlangen, Germany

[2] Advanced Light Source, Lawrence Berkeley National Laboratory, Berkeley, CA, USA

[3] Department of Molecular Physics, Fritz-Haber-Institut der Max-Planck-Gesellschaft, Berlin, Germany

[4] Lehrstuhl für Angewandte Physik, Friedrich-Alexander-Universität Erlangen-Nürnberg, Erlangen, Germany

[5] Sandia National Laboratories, Albuquerque, NM, USA

[6] National Center for Electron Microscopy, Lawrence Berkeley National Laboratory, Berkeley, CA, USA

* Corresponding author: thomas.seyller@physik.uni-erlangen.de

**Abstract**

We have investigated epitaxial graphene films grown on SiC(0001) by annealing in an atmosphere of Ar instead of vacuum. Using AFM and LEEM we observe a significantly improved surface morphology and graphene domain size. Hall measurements on monolayer graphene films show a carrier mobility of around 1000 $cm^2$/Vs at room temperature and 2000 $cm^2$/Vs at 27K. The growth process introduced here establishes the synthesis of graphene films on a technologically viable basis.

**Introduction**

Graphene, a single monolayer of graphite, has recently attracted considerable interest due to its novel magneto transport properties [1-3], high carrier mobility, and ballistic transport up to room temperature [4]. It has technological applications as a potential successor of silicon for the post Moore's law era [5-7], as single molecule gas sensor[8], in spintronics [9-11], in quantum computing [12], or as terahertz oscillators [13]. For such applications, uniform ordered growth of graphene on an insulating substrate is necessary. While on the one hand exfoliation of graphene from graphite yields high quality crystals, such isolated samples with dimensions in the 10 micrometer range are unsuitable for large-scale device production; on the other hand the vacuum decomposition of SiC yields wafer-sized samples with small grains (30-200 nm) that are equally unsuitable. Here we show that the *ex-situ* graphitization of Si-terminated SiC(0001) in an argon atmosphere of about 1 bar produces monolayer graphene films with much larger domain sizes than previously attainable. Hall measurements confirm the quality of the films thus obtained. For two different geometries, high electronic mobilities were found, which reach $\mu$ = 2000 cm$^2$/Vs at $T$=27 K. The new growth process introduced here establishes the synthesis of graphene films on a technologically viable basis.

The preparation of single layer graphene by the thermal decomposition of SiC is envisaged as a viable route for the synthesis of uniform, wafer-size graphene layers for technological applications, but the large scale structural quality is presently limited by the lack of continuity and uniformity of the grown film [14, 15]. On the Si-terminated (0001) basal plane, vacuum annealing leads to small graphene domains typically 30-100 nm in diameter, while on the C-terminated ($000\bar{1}$) face, larger domains (~200 nm) of multilayered, rotationally disordered graphene have been produced [16]. The small-grain structure is due to morphological changes of the surface in the course of high temperature annealing. Moreover, decomposition of SiC is not a self-limiting process and, as a result, regions of different film thicknesses coexist, as shown by low-energy electron microscopy (LEEM) [14, 15]. Such inhomogeneous films do not meet the demands of large scale device production which requires larger grains and tighter thickness control. Homogeneous film thickness is particularly important because the electronic structure of the film depends strongly on the number of layers. For example, while monolayer graphene is a gapless semiconductor, a forbidden gap can be induced in bilayer graphene and tuned by an external electrostatic potential [12, 17-20].

**Experimental**

Graphene layers were synthesized on commercial, nominally on-axis oriented wafers of 6H-SiC(0001) purchased from SiCrystal AG. Prior to graphene epitaxy the samples were etched in hydrogen (grade 5.0, $p$ = 1 bar, $T$ = 1550°C, $t$ = 15 minutes) in order to remove surface polishing damage. Graphene growth was carried out in a vertical cold wall reactor comprised of a double-walled, water-cooled quartz tube and a graphite susceptor in a slow flow of argon (purity 5.0). Heating and cooling rates were 2-3°C per second. Typical annealing time was 15 minutes. A wide range of annealing temperatures from 1500 to 2000° C and reactor gas pressures from 10 mbar to 900 mbar were tested and a detailed account of all observations will be provided elsewhere. However, except for very low pressures studied the morphology of the surface after graphene formation in Ar atmosphere is generally much smoother and the graphene domain size much larger compared to vacuum annealing.

Surface composition and graphene thicknesses were obtained from core-level photoelectron spectroscopy (XPS) by means of a Specs PHOIBOS150 analyzer in combination with a monochromatized Al $K_\alpha$ source with an energy resolution of ~350 meV. Owing to the chemical inertness of graphene the samples can be easily transported through air. As-prepared graphene samples showed no detectable oxygen on the surface (below 1% of a monolayer) even after air exposure for about 1 hour. Prolonged air exposure, however, leads to a fractional layer of physisorbed hydrocarbons and water which can be removed by annealing in vacuum at around 600°C. ARPES measurements were performed at the Advanced Light Source (ALS) using a Scienta R4000 analyzer with an overall resolution of ~25 meV at a photon energy of 94 eV and at a sample temperature of 20 K. Core level measurements were performed at BESSY-II with a Specs PHOIBOS150 analyzer with a resolution of ~125 meV at a photon energy of 700eV and at room temperature. The surface morphology was probed by atomic force microscopy (AFM) in non-contact mode. LEEM measurements at room temperature with a spatial resolution better than 10 nm were carried out at Sandia National Laboratory and at the National Center for Electron Microscopy, Lawrence Berkeley National Laboratory. The crystal structure of the films was monitored by low-energy electron diffraction (LEED).

For the electrical characterization, the samples were patterned by two electron beam lithography steps: The first step defined the graphene film (undesired areas were etched with oxygen plasma). A second step defined the contact pads, which consist of thermally evaporated Ti/Au double layer, patterned by a standard lift-off technique. Electrical measurements in van der Pauw geometry or on Hall bar structures were carried out in a continuous flow cryostat (sample in vacuum), using magnetic fields of ±0.66 T at temperatures between 300 and 25 K.

**Results and discussion**

Consider the data in Fig. 1, where we compare samples prepared by vacuum annealing with samples produced by *ex-situ* annealing under argon atmosphere. Panels (a)-(c) show the morphology of the SiC (0001) surface before and after the formation of a graphene monolayer by annealing in ultra high vacuum (UHV) as determined by atomic force microscopy (AFM) and LEEM. The initial SiC(0001) surface in Fig.1(a), obtained after hydrogen etching, is characterized by wide, highly uniform, atomically flat terraces. The step direction and terrace width (on the order of 300-700 nm) are determined by the incidental misorientation of the substrate surface with respect to the crystallographic (0001) plane. The step height is 15 Å which corresponds to the dimension of the 6H-SiC unit cell in the direction perpendicular to the surface (c-axis). On defect-free areas of the sample, the terraces typically extend undisturbed over 50 μm in length. The morphology of the surface covered with a monolayer of graphene prepared by vacuum annealing is shown in Fig. 1(b). The surface obviously undergoes significant modifications; it is now covered with small pits up to 10 nm in depth, and the original steps are hardly discernible any longer. This indicates that graphene growth is accompanied by substantial changes in the morphology of the substrate itself, leading to a considerable roughening. As a consequence of this roughening, the graphene layer acquires an inhomogeneous thickness distribution as can be seen in the LEEM image shown in Fig.1(c). The irregularly-shaped graphene islands are at most a few hundred nm in size, in agreement with x-ray diffraction [16]. Moreover, monolayer graphene areas coexist with graphene bilayer islands as well as with uncovered regions of the (6√3×6√3) buffer layer[21].

In stark contrast to the low quality resulting from vacuum graphitization (Fig. 1(b)), films grown under 900 mbar of argon have a greatly improved surface morphology, as demonstrated by the AFM image in Fig. 1(d). Step bunching is manifested by the formation of macro-terraces with a width that increases from about 0.5 μm on the original surface (Fig. 1(a)) to about 3 μm. Correspondingly, the macro-steps which are running in the same crystallographic direction as the original steps reach a height of about 15 nm. Parallel to the steps, uninterrupted macroterraces more than 50 μm long have been observed.

The thickness distribution of the graphene film grown *ex-situ* under an argon atmosphere is determined by LEEM as shown in figs. 1(e,f). Series of spatially-resolved LEEM I-V spectra taken along a vertical and a horizontal line in fig. 1(f) are shown in figs. 1(g,h). The layer thickness is easily determined from the number of minima in the individual spectra; the LEEM image taken at a particular energy shows stripes that follow in width and orientation the macroterraces with a contrast that is determined by the graphene layer thickness. [14, 15] Hence, we can unambiguously conclude that except for narrow stripes at the edges, the large atomically flat macro-terraces are homogeneously covered with a graphene monolayer. The domain size of monolayer graphene is significantly larger than that of the vacuum annealed samples as a comparison between figs. 1(c) and 1(f) shows. In fact, the domain size appears to be limited by the length and width of the SiC terraces only. Narrower, darker regions at the downward edges of the terraces correspond to bilayer and in some cases trilayer graphene (see region 3 in fig. 1(f)). In the AFM image these regions (see fig. 1(i)) appear as small depressions of around 4 Å and 8 Å amplitude located at the very edge of the macrostep. This indicates that the nucleation of new graphene layers starts at step edges of the substrate surface. We also note that the laterally averaged graphene thickness determined by LEEM is in perfect agreement with the average layer thickness value of 1.2 ML obtained by x-ray photoelectron spectroscopy (XPS).

The structural and electronic properties of graphene layers grown under an argon atmosphere were also probed by low-energy electron diffraction (LEED) and photoelectron spectroscopy, as shown in Figure 2. While these methods alone cannot assess the morphological quality of the sample surface on a large scale, they provide additional information as detailed below. Firstly, the LEED pattern (Fig. 2(a)) demonstrates that the graphene layer is well ordered and aligned with respect to the substrate, such that the basal plane unit vectors of graphene and SiC subtend an angle of 30 degrees. The C1s core level spectrum (Fig. 2(b)) shows the characteristic signals of the SiC substrate, the (6√3×6√3) interface layer and the graphene monolayer, respectively, in excellent agreement with previous work [21, 22]. Finally, the angle-resolved photoelectron spectroscopy (ARPES) measurement (Fig. 2(c)) reveals the characteristic band structure of monolayer graphene [20, 23, 24]. Note that, as for vacuum grown layers [20, 23, 24], the Dirac point ($E_D$) is shifted below the Fermi level ($E_F$) due to electron doping ($n \approx 1.1 \times 10^{13} cm^{-2}$) from the substrate. Therefore, while our epitaxial growth process results in a dramatic improvement in surface morphology all other important properties such as crystalline order, electronic structure, and charge carrier density remain unaltered as compared to vacuum grown layers.

What is the reason for the observed improvement of the surface morphology of the Ar-annealed samples compared to the samples annealed in UHV? From the data in Fig. 1. it is clear that the surface undergoes considerable morphological changes at the temperature where graphitization occurs. The large roughness of the UHV annealed samples suggests that the surface is far from equilibrium, such that a transformation to a smooth morphology cannot be achieved under these conditions. The key factor in achieving an improved growth is the *significantly higher annealing temperature* of 1650°C that is required for graphene formation under argon at a pressure of 900 mbar as compared to 1280°C in UHV. Graphene formation is the result of Si evaporation from the substrate. For a given temperature, the presence of a high pressure of argon leads to a reduced Si evaporation rate because the silicon atoms desorbing from the surface have a finite probability of being reflected back to the surface by collision with Ar atoms, as originally pointed out by Langmuir [25, 26]. Indeed, in presence of the Ar atmosphere no sublimation of Si from the surface is observed at temperatures up to 1500°C whereas Si desorption commences at 1150°C in vacuum. The significantly higher growth temperature thus attained results in an enhancement of surface diffusion such that the restructuring of the surface that lowers the surface free energy (by step bunching, for example) is completed before graphene is formed. Ultimately, this leads to the dramatically improved surface morphology that we observe here. The macrostep structure is also responsible for the tighter thickness control. As shown above, a new graphene layer starts to grow from the step edges; hence having fewer steps along well defined crystallographic directions reduces the nucleation density of multilayer graphene.

In order to evaluate the electronic quality of our graphene layers we determined the carrier mobility of monolayer epitaxial graphene on SiC(0001) using Hall effect measurements. Two different geometries were investigated, both patterned with electron beam lithography: square graphene films (100 μm × 100 μm) with contact pads at the four corners for van der Pauw measurements as well as Hall bars (2 μm × 30 μm) placed on macroterraces. No significant difference in electron mobility was observed between the two geometries indicating that step edges play a minor role. Mobilities of 930 cm$^2$/Vs and 2000 cm$^2$/Vs were measured at 300 and 27 K, respectively. At the same time the electron density remained basically constant $n \approx 1.1 \times 10^{13}$ cm$^{-2}$. Therefore we should compare our mobility with exfoliated monolayer graphene on SiO$_2$ in the high doping limit, where values of the order of 10000 cm$^2$/Vs are reported for $n \approx 5 \times 10^{12}$ cm$^{-2}$ [2, 27]. For epitaxial graphene on SiC, other groups reported on maximum values of 1200 cm$^2$/Vs for *multilayered* graphene [6, 28].

Figure 3 shows the temperature dependence of the electron mobility measured in van der Pauw geometry. The linear $\mu(T)$ dependence is unexpected. Scattering at acoustic phonons of graphene would result in $T^4$ behavior at low temperatures [29]. A theoretical treatment of the effect of static impurities in graphene predicts 1/T dependence of the mobility [30]. Candidates for such impurities are certainly dangling bonds below the graphene layer. Also adsorbates might play a certain role [8]. The linear dependence of the scattering rates rather fits to the case of electron-electron interaction in a 2D electron gas [31]. Clearly more work is required to understand the temperature dependence of the mobility in epitaxial graphene.

**Conclusion**

In conclusion, we have shown that the growth of epitaxial graphene on SiC(0001) in an Ar atmosphere close to atmospheric pressure provides morphologically superior graphene layers in comparison to vacuum graphitization. Extensive step bunching taking place during processing yields arrays of parallel terraces up to 3 µm wide and more than 50 µm long. The terraces are essentially completely and homogeneously covered with a monolayer of graphene. At present, downward step edges, where the initiation of second and third layer graphene growth is detected, are prohibiting an even larger extend of the graphene domains. Because the substrate step direction and step width are determined by the magnitude and azimuthal orientation of the surface misorientation with respect to major crystallographic directions, a proper choice of these parameters controls terrace width and length and hence the ultimate uninterrupted lateral extent of the graphene layer. An improved substrate quality in terms of crystallographic orientation is therefore expected to lead to further improvements. In comparison to the UHV treatment, the technique presented here is much closer to standard preparation conditions in semiconductor manufacture, permitting the use of standard CVD (chemical vapor deposition) equipment for the fabrication of graphene layers. All necessary processing steps, i.e. hydrogen etching and graphene synthesis, can be carried out in a single reactor. Electrical measurements on monolayer graphene layers confirm the picture of improved film quality: mobilities around 1000 cm$^2$/Vs at room temperature, which increases linearly up to 2000 cm$^2$/Vs at 27 K.

**Acknowledgements**

We thank F. El Gabaly for assistance with the LEEM measurements. We gratefully acknowledge support by the DFG under contract SE 1087/5-1, contract WE4542/5-1, and within the Cluster of Excellence 'Engineering of Advanced Materials' (www.eam.uni-erlangen.de) at the Friedrich-Alexander-Universität Erlangen-Nürnberg, and the BMBF under contract 05 ES3XBA/5. A part of the work was performed at Sandia National Laboratories, a multiprogram laboratory operated by Sandia Corporation, a Lockheed Martin company, for the United States Department of Energy, Office of Basic Energy Sciences, Division of Materials Science and Engineering, under Contract No. DE-AC04-94AL85000. The work performed at the ALS supported by the Director, Office of Science, Office of Basic Energy Sciences, of the U.S. Department of Energy under Contract No. DE-AC03-76SF00098.

**References**

[1]     K. S. Novoselov *et al.*, Nature **438**, 197 (2005).
[2]     Y. B. Zhang *et al.*, Nature **438**, 201 (2005).
[3]     K. S. Novoselov *et al.*, Nat. Phys. **2**, 177 (2006).
[4]     K. S. Novoselov *et al.*, Science, 1137201 (2007).
[5]     A. K. Geim, and K. S. Novoselov, Nat. Mater. **6**, 183 (2007).
[6]     C. Berger *et al.*, J. Phys. Chem. B **108**, 19912 (2004).
[7]     C. Berger *et al.*, Science **312**, 1191 (2006).
[8]     F. Schedin *et al.*, Nat. Mater. **6**, 652 (2007).
[9]     Y.-W. Son, M. L. Cohen, and S. G. Louie, Nature **444**, 347 (2006).
[10]    B. Trauzettel *et al.*, Nat. Phys. **3**, 192 (2007).
[11]    T. Yokoyama, Phys. Rev. B **77**, 073413 (2008).
[12]    V. I. Fal'ko, Nat. Phys. **3**, 151 (2007).
[13]    F. Rana, IEEE Trans. Nanotechnol. **7**, 91 (2008).
[14]    H. Hibino *et al.*, Phys. Rev. B **77**, 075413 (2008).
[15]    T. Ohta *et al.*, New J. Phys. **10**, 023034 (2008).
[16]    J. Hass *et al.*, Appl. Phys. Lett. **89**, 143106 (2006).
[17]    F. Guinea, A. H. C. Neto, and N. M. R. Peres, Phys. Rev. B **73**, 245426 (2006).
[18]    E. McCann, Phys. Rev. B **74**, 161403 (2006).
[19]    E. McCann, and V. I. Fal'ko, Phys. Rev. Lett. **96**, 086805 (2006).
[20]    T. Ohta *et al.*, Science **313**, 951 (2006).
[21]    K. V. Emtsev *et al.*, Phys. Rev. B **77**, 155303 (2008).
[22]    K. V. Emtsev *et al.*, Mater. Sci. Forum **556-557**, 525 (2007).
[23]    A. Bostwick *et al.*, Nat. Phys. **3**, 36 (2007).
[24]    T. Ohta *et al.*, Phys. Rev. Lett. **98**, 206802 (2007).
[25]    I. Langmuir, Phys. Rev. (Series I) **34**, 401 (1912).
[26]    G. R. Fonda, Phys. Rev. (Series II) **31**, 260 (1928).
[27]    K. S. Novoselov *et al.*, Science **306**, 666 (2004).
[28]    J. Kedzierski *et al.*, IEEE Transactions on Electron Devices **55**, 2078 (2008).
[29]    E. H. Hwang, and S. D. Sarma, Phys. Rev. B **77**, 115449 (2008).
[30]    V. V. Cheianov, and V. I. Fal'ko, Phys. Rev. Lett. **97**, 226801 (2006).
[31]    G. Zala, B. N. Narozhny, and I. L. Aleiner, Phys. Rev. B **64**, 214204 (2001).

**Figure captions**

Fig 1: Morphological changes of 6H-SiC(0001) during graphene growth. (a) Initial surface after H-etching imaged by AFM. The step height is 15 Å. (b) AFM image of graphene on 6H-SiC(0001) with a nominal thickness of 1 ML formed by annealing in UHV at a temperature of about 1280°C. (c) LEEM image of a UHV grown graphene film on SiC(0001) with a nominal thickness of 1.2 monolayers. The image contrast is due to the locally different layer thickness. Light, medium, and dark gray correspond to a local thickness of 0, 1, and 2 ML, respectively. (d) AFM image of graphene on 6H-SiC(0001) with a nominal thickness of 1.2 ML formed by annealing in Ar (p=900 mbar, T= 1650°C). (e) LEEM image of a sample equivalent to that of (d) revealing macroterraces covered with graphene up to 50µm long and at least 1µm wide. (f) Close-up LEEM image revealing monolayer coverage on the terraces and bilayer/trilayer growth at the step edges. (g,h) Electron reflectivity spectra (gray scale images) taken at positions indicated by the blue lines in (f). Monolayer, bilayer, and trilayer graphene are readily identified by the presence of 1, 2, or 3 reflectivity minima, respectively. (i) Close-up AFM images of the film shown in (d). In the right hand side image the z-scale was adjusted such that the terraces appear at the same height. The profile shows that small depressions 4 and 8 Å in height exist at the step edges due to $2^{nd}$ and $3^{rd}$ layer nucleation.

Fig 2. Atomic and electronic structure of *ex-situ* grown monolayer graphene. (a) LEED pattern at 74 eV showing the diffraction spots due to the SiC(0001) substrate (blue arrows) and the graphene lattice (red arrows). The additional spots are due to the (6√3×6√3) interface layer. (b) C1s core level spectrum measured at a photon energy of 700 eV. The spectrum contains contributions from the SiC substrate (marked SiC), the (6√3×6√3) interface layer (marked S1 and S2), and from the graphene layer (G) residing on top of the interface layer. (c) π-bands probed by ARPES in the vicinity of the K-point of the hexagonal Brillouin zone measured along ΓK direction. The position of the Dirac Energy ($E_D$) at 0.45 eV below the Fermi energy is consistent with previous reports on UHV grown graphene on SiC(0001). Faint features marked by yellow arrows signal the presence of small regions of bilayer graphene in agreement with the LEEM results.

Fig. 3. Temperature dependence of electron mobility in a monolayer epitaxial graphene. The mobility values were derived from Hall measurements on a sample in van der Pauw geometry. The experimental data display a linear *T* dependence.

Fig. 1

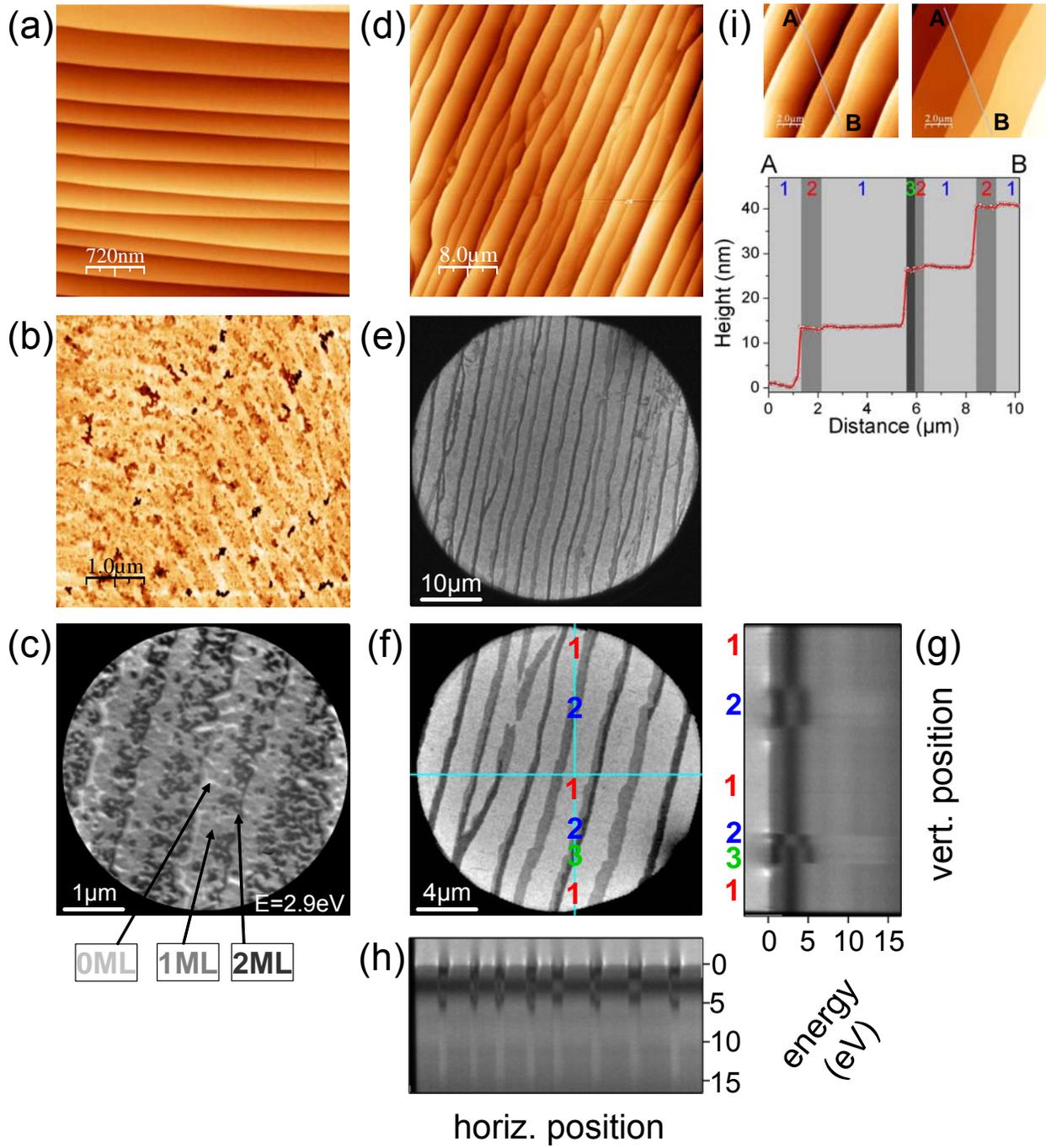

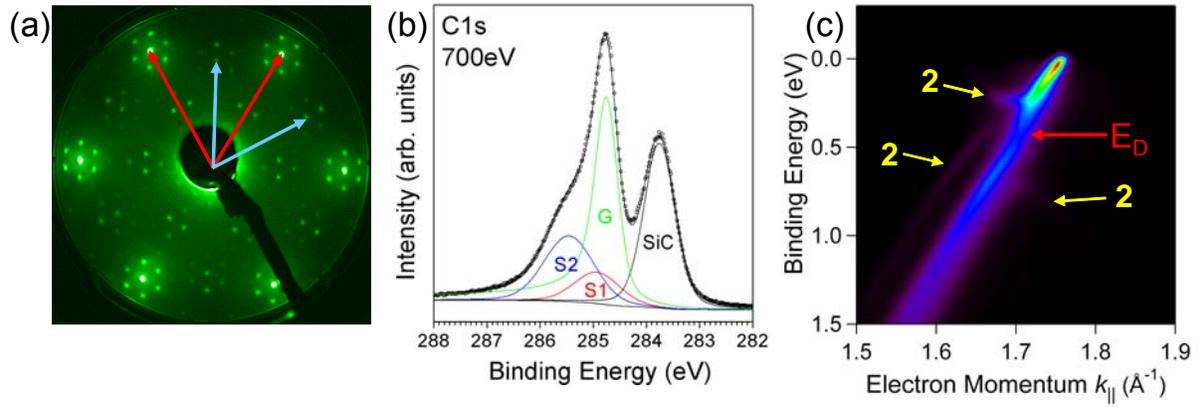

**Fig. 3**

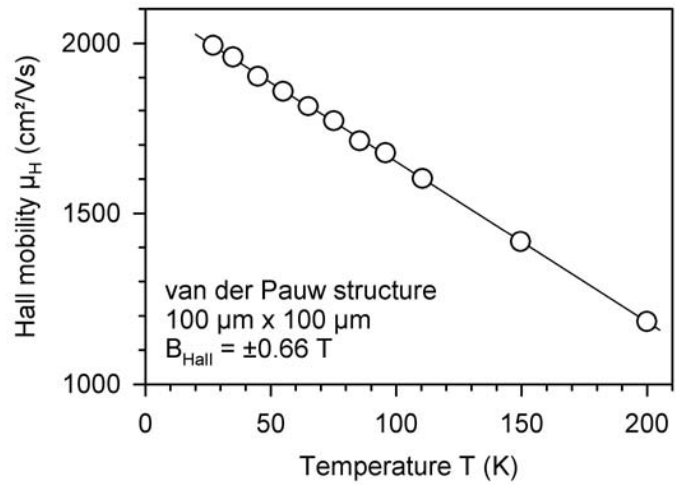